\documentclass{article}

\usepackage[preprint]{neurips_2026}

\usepackage[utf8]{inputenc}
\usepackage[T1]{fontenc}
\usepackage{hyperref}
\usepackage{url}
\usepackage{booktabs}
\usepackage{amsfonts}
\usepackage{amsmath}
\usepackage{nicefrac}
\usepackage{microtype}
\usepackage{xcolor}
\usepackage{graphicx}
\usepackage{listings}
\usepackage{array}
\usepackage{multirow}
\usepackage{float}

\newcommand{\shape}[1]{\texttt{(#1)}}
\newcommand{\block}[1]{\textbf{Block~#1}}
\newcommand{\wavlm}{\textsc{WavLM}}

\title{DiariZen Explained: A Tutorial for the Open Source State-of-the-Art Speaker Diarization Pipeline}

\author{%
  Nikhil Raghav \\
  Institute for Advancing Intelligence, TCG CREST, Kolkata -- 700\,091, India \\Department of Computer Science, RKMVERI, Howrah -- 711 202, India \\
  \texttt{nikhil.raghav.92@tcgcrest.org}
}

\begin{document}

\maketitle

\begin{abstract}
Speaker diarization (SD) is the task of answering \emph{who spoke when} in a multi-speaker audio stream. Classically, an SD system clusters segments of speech belonging to an individual speaker's identity. Recent years have seen substantial progress in SD through end-to-end neural diarization (EEND) approaches. DiariZen~\citep{han2025leveraging}, a hybrid SD pipeline built upon a structurally pruned WavLM-Large encoder, a Conformer backend with powerset classification, and VBx clustering, represents the leading open-source state of the art at the time of writing across multiple benchmarks. Despite its strong performance, the DiariZen architecture spans several repositories and frameworks, making it difficult for researchers and practitioners to understand, reproduce, or extend the system as a whole. This tutorial paper provides a self-contained, block-by-block explanation of the complete DiariZen pipeline, decomposing it into
seven stages: (1) audio loading and sliding window segmentation, (2) WavLM feature extraction with learned layer weighting, (3) Conformer backend and powerset classification, (4) segmentation aggregation via overlap-add, (5) speaker embedding extraction with overlap exclusion, (6) VBx clustering with PLDA scoring, and (7) reconstruction and RTTM output. For each block, we provide the conceptual motivation, source code references, intermediate tensor shapes, and annotated visualizations of the actual outputs on a 30\,s excerpt from the AMI Meeting Corpus. The implementation is available at \url{https://github.com/nikhilraghav29/diarizen-tutorial}, which includes standalone executable scripts for each block and a Jupyter notebook that runs the complete pipeline end-to-end.
\end{abstract}

\section{Introduction}
\label{sec:intro}

Speaker diarization is one of the foundational tasks at the intersection of conversational AI and audio analytics. Given an audio recording containing multiple (at least two) speakers, a diarization system must produce a time-annotated speaker segmentation of the form ``Speaker A spoke from 0.8\,s to 13.6\,s, Speaker B spoke from 13.7\,s to 18.4\,s,'' and so forth. It has broad implications in many downstream tasks, including automatic speech recognition (ASR) in meetings, audio forensics, clinical documentation, and media analytics~\citep{park2022review}.

Classical diarization pipelines are modular, spanning across several stages including voice activity detection, segmentation, speaker embedding extraction, and clustering, which are performed as independent sequential steps. While modular, these systems suffer from cascading errors; mistakes in early stages cannot be corrected by later ones. The emergence of EEND approaches~\citep{fujita2019eend} addressed this by training the full pipeline jointly, but early EEND models were limited to a fixed maximum number of speakers and struggled with long recordings.

DiariZen~\citep{han2025leveraging, han2025fine, han2025efficient} represents a principled synthesis of both paradigms. It combines an EEND-style neural segmentation front-end based on a structurally pruned WavLM-Large encoder and a Conformer with powerset classification with a VBx clustering back-end that resolves global speaker identities across overlapping windows. The system currently achieves state-of-the-art performance on multiple benchmarks, including AMI, VoxSRC, and DIHARD-III.

\paragraph{Motivation for this tutorial.}
Despite DiariZen's strong performance, its implementation is distributed across three separate repositories: the main DiariZen~\footnote{\url{https://github.com/BUTSpeechFIT/DiariZen}}~\citep{han2025leveraging} repository, a modified fork of \texttt{pyannote-audio}~\footnote{\url{https://github.com/BUTSpeechFIT/DiariZen/tree/main/pyannote-audio}}~\citep{bredin2023pyannote}, and a custom \texttt{torchaudio}-based WavLM implementation supporting structured pruning through knowledge distillation~\citep{peng2023dphubert}. There is no single accessible resource that explains the system from raw audio input to rich transcription time marked (RTTM) output, traces intermediate tensor shapes through every transformation, or provides executable code for each individual stage.

This tutorial fills this gap through the following contributions:

\begin{itemize}
  \item A self-contained block-by-block explanation of the complete
    DiariZen pipeline with annotated tensor shapes at every stage.
  \item Annotated visualizations of intermediate outputs on a real
    multi-speaker recording, including WavLM layer weights, powerset
    class probabilities, overlap-add coverage maps, embedding cosine
    similarity matrices, and VBx cluster assignments.
  \item A modular GitHub repository with standalone Python scripts for each block, a \texttt{pipeline\_loader.py} utility, and a Jupyter notebook for interactive exploration. 

\end{itemize}

The remainder of the tutorial is organized as follows. Section~\ref{sec:background} provides the background necessary for familiarization with the main ideas used in the pipeline.
Section~\ref{sec:overview} gives a system-level overview of the main blocks introduced in the pipeline.
Sections~\ref{sec:block1}--\ref{sec:block7} explain each block in
detail. Section~\ref{sec:experiment} presents experimental results on
the sample recording from the AMI corpus. 
Section~\ref{sec:conclusion} concludes the tutorial and outlines the potential future research directions.

\section{Background}
\label{sec:background}

\subsection{What is Speaker diarization?}

Given a single-channel audio waveform $\mathbf{x} \in \mathbb{R}^T$
sampled at 16\,kHz, a diarization system must produce a set of labelled
time intervals $\{(t_s^{(i)}, t_e^{(i)}, \ell^{(i)})\}_{i=1}^N$, where
$t_s^{(i)}$ and $t_e^{(i)}$ are the start and end times of the $i$-th
segment and $\ell^{(i)} \in \{\text{SPK\_00}, \text{SPK\_01}, \ldots\}$
is the speaker label. 

The performance of the diarization system is measured through the standard metric diarization error
rate~(DER):
\begin{equation}
  \text{DER} = \frac{T_{\text{miss}} + T_{\text{fa}} +
  T_{\text{conf}}}{T_{\text{ref}}},
\end{equation}
where $T_{\text{miss}}$, $T_{\text{fa}}$, and $T_{\text{conf}}$ denote
the total duration of missed speech, false alarm speech, and the speaker
confusion or misclassifications errors respectively, and $T_{\text{ref}}$ is the total
reference speech duration.

\subsection{EEND-VC architecture}

End-to-end neural diarization with vector clustering
(EEND-VC)~\citep{kinoshita2021integrating} is a hybrid paradigm that combines the best of both worlds: classical and EEND models. It combines a local neural
segmentation model with a global clustering step. The neural model
processes short overlapping windows independently and assigns local
speaker activity scores on the frame level. A clustering algorithm then resolves and maps 
the local speaker labels across windows into consistent global
identities. This hybrid approach handles arbitrarily long recordings and
an unknown number of speakers, unlike pure EEND models that are
constrained to a fixed speaker count.

\subsection{Powerset encoding}

Conventionally, EEND models predict speaker activity independently per speaker: a multi-label classification where each output neuron corresponds to one speaker. Powerset encoding~\citep{plaquet2023powerset}, by contrast, predicts the \emph{combination} of active speakers as a single categorical label. The name derives directly from the concept of a powerset in set theory: the set of all subsets of a given set. 

Given a maximum of $S$ speakers per chunk and a
maximum of $O$ overlapping speakers per frame, the number of powerset
classes is:

\begin{equation}
  K = \sum_{k=0}^{O} \binom{S}{k}.
\end{equation}
In the DiariZen configuration, $S=4$ and $O=2$. 

In this case, $K = \binom{4}{0} + \binom{4}{1} + \binom{4}{2} = 1 + 4 + 6 = 11$ classes.
Explicitly, the powerset $\mathcal{P}$ over the set of local speakers
$\mathcal{S} = \{s_1, s_2, s_3, s_4\}$, restricted to subsets of
cardinality at most $O = 2$, is:

\begin{equation}
\mathcal{P} = \bigl\{
  \underbrace{\emptyset}_{\text{silence}},\;
  \underbrace{\{s_1\},\, \{s_2\},\, \{s_3\},\, \{s_4\}}_{\text{single speaker }(4)},\;
  \underbrace{\{s_1,s_2\},\, \{s_1,s_3\},\, \{s_1,s_4\},\,
              \{s_2,s_3\},\, \{s_2,s_4\},\, \{s_3,s_4\}}_{\text{two-speaker overlap }(6)}
\bigr\},
\end{equation}

\noindent giving $|\mathcal{P}| = 1 + 4 + 6 = 11$ classes in total.
Each frame of the audio is assigned exactly one element of $\mathcal{P}$,
representing the set of speakers simultaneously active at that instant.
The conventional multi-label formulation instead assigns each speaker
an independent binary label, producing $2^4 = 16$ possible
combinations~--- including the four three-speaker and one four-speaker
combinations that the powerset formulation deliberately excludes by
enforcing $O = 2$.
\subsection{WavLM \& self-supervised speech representations}

\wavlm~\citep{chen2022wavlm} is a self-supervised speech foundation
model based on the masked speech prediction framework of
HuBERT~\citep{hsu2021hubert}, extended with denoising objectives. It is
pretrained on 94,000 hours of diverse audio and produces rich
frame-level representations across 24 transformer layers. Following the
SUPERB benchmark~\citep{yang2021superb}, downstream tasks combine all
layers via a learned scalar weighted sum rather than using only the
final layer. DiariZen uses a structurally pruned variant (80\%
parameter reduction, 316M $\rightarrow$ 63M parameters) obtained
through DPHuBERT-style pruning~\citep{peng2023dphubert}.

\subsection{VBx clustering}

Variational Bayes x-vector clustering
(VBx)~\citep{diez2019bayesian} is a probabilistic clustering
algorithm for speaker diarization that operates in two stages. First,
agglomerative hierarchical clustering (AHC) with probabilistic linear discriminant analysis
(PLDA) scoring~\citep{prince2007probabilistic} produces an initial
speaker assignment. Second, a Variational Bayes expectation-maximisation
procedure over a hidden Markov model (VB-HMM) refines the assignments
iteratively. VBx is parameter-efficient, requiring only a small PLDA
model, and produces globally consistent speaker identities across
arbitrarily long recordings.

\section{System Overview}
\label{sec:overview}

\begin{figure}[h]
  \centering
  \includegraphics[width=1.0\linewidth]{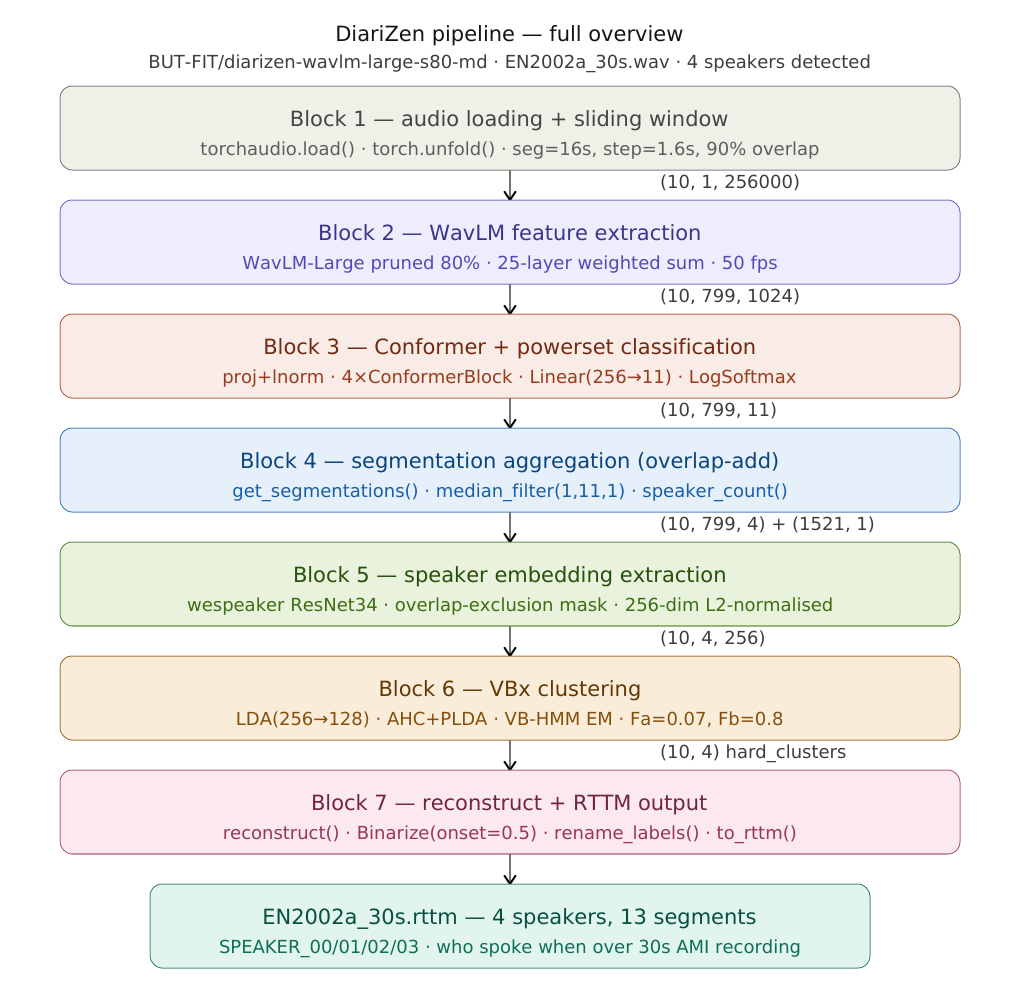}
\caption{The sequential blocks illustrating the complete DiariZen
pipeline. A 30\,s multi-speaker recording is first segmented into 10
overlapping 16\,s chunks using a sliding window with 90\% overlap
(Block~1). Each chunk is passed through the pruned WavLM-Large
encoder to extract 1024-dim frame-level features via a learned
layer-weighted sum (Block~2), which are then processed by a
4-layer Conformer backend to produce per-frame powerset
segmentation scores over 11 speaker-combination classes (Block~3).
The per-chunk scores are aggregated across overlapping windows via
overlap-add and median filtering to yield a continuous segmentation
and instantaneous speaker count (Block~4). Speaker embeddings are
then extracted per (chunk, speaker) pair using the WeSpeaker
ResNet34 model with overlap exclusion masking (Block~5), and
clustered into global speaker identities via VBx with PLDA
scoring and VB-HMM refinement (Block~6). Finally, the cluster
assignments are mapped back to a continuous diarization annotation
and written as an RTTM file (Block~7).}
  \label{fig:overview}
\end{figure}

Figure~\ref{fig:overview} shows the complete DiariZen pipeline, mentioning all the main components in each block, with their respective choice of parameters. The SD
system accepts a raw audio file as input and produces an RTTM diarization
annotation through seven sequential blocks. Table~\ref{tab:shapes}
summarizes the tensor shape at the output of each block for the
30-second AMI test recording used throughout this tutorial.

\begin{table}[h]
  \caption{Tensor shape at the output of each block for the AMI
    \texttt{EN2002a\_30s.wav} test recording (30\,s, 16\,kHz, mono).}
  \label{tab:shapes}
  \centering
  \begin{tabular}{clll}
    \toprule
    Block & Name & Output shape & Description \\
    \midrule
    1 & Sliding window     & \shape{10, 1, 256000}  & 10 chunks $\times$ 1 channel $\times$ 256{,}000 samples \\
    2 & WavLM features     & \shape{10, 799, 1024}  & 10 chunks $\times$ 799 frames $\times$ 1024-dim \\
    3 & Powerset scores    & \shape{10, 799, 11}    & log-softmax over 11 speaker combinations \\
    4 & Segmentation (OLA) & \shape{10, 799, 4}     & per-speaker binary activity + \shape{1521,1} count \\
    5 & Embeddings         & \shape{10, 4, 256}     & 256-dim L2-normalised per (chunk, speaker) \\
    6 & Hard clusters      & \shape{10, 4}          & global speaker ID, $-2$ = inactive \\
    7 & RTTM output        & 13 segments            & 4 global speakers, \texttt{.rttm} file \\
    \bottomrule
  \end{tabular}
\end{table}

\paragraph{Hardware and software requirements.}
All experiments were run on an NVIDIA H200~NVL GPU (150\,GB VRAM).
The software environment uses Python~3.9, PyTorch~2.1.2, and the
\texttt{diarizen} conda environment whose specification is provided in
the repository. The pretrained model
\texttt{BUT-FIT/diarizen-wavlm-large-s80-md}~\footnote{\url{https://huggingface.co/BUT-FIT/diarizen-wavlm-large-s80-md}} (278\,MB) and the
WeSpeaker embedding model
\texttt{pyannote/wespeaker-voxceleb-resnet34-LM}~\footnote{\url{https://huggingface.co/pyannote/wespeaker-voxceleb-resnet34-LM}} (27\,MB) are loaded
from the HuggingFace Hub.

\section{Block 1 --- Audio Loading and Sliding Window Segmentation}
\label{sec:block1}

\paragraph{Concept:} The first block converts a raw WAV file into a batch of overlapping
fixed-length audio chunks. This windowing strategy is fundamental to
EEND-VC: the neural segmentation model was trained on short, fixed-length
segments, so inference must respect the same duration at test time.
Overlap between consecutive windows ensures that every frame of the
recording is covered by multiple independent predictions, which are
later averaged in \block{4} to reduce noise.

\paragraph{Source code:} This block corresponds to \texttt{DiariZenPipeline.\_\_call\_\_()} in
\texttt{diarizen/pipelines/inference.py}, which calls
\texttt{torchaudio.load()} followed by
\texttt{pyannote.audio.core.Inference.slide()}, which internally applies \texttt{torch.Tensor.unfold()} for windowing.

\paragraph{Implementation:}
Given a single-channel waveform $\mathbf{x} \in \mathbb{R}^T$ at 16\,kHz, with
$T = 480{,}000$ samples (30 seconds), the sliding window produces chunks
of size $W = \text{seg\_duration} \times f_s = 16 \times 16{,}000 =
256{,}000$ samples with step $H = \text{segmentation\_step} \times W
= 0.1 \times 256{,}000 = 25{,}600$ samples, corresponding to a 90\%
overlap. 

The number of complete chunks is:
\begin{equation}
  N_c = \left\lfloor \frac{T - W}{H} \right\rfloor + 1 = 9,
\end{equation}
with one additional zero-padded chunk covering the final
$T - (N_c-1)\cdot H = 249{,}600$ samples (15.6\,s of real audio and
0.4\,s of silence). The output is a tensor of shape \shape{10, 1,
256000}.

\paragraph{Key design choices:} The 90\% overlap (step ratio 0.1) ensures that overlap-add
aggregation in \block{4} averages up to 10 independent predictions for
every interior frame of the recording. This heavy overlap substantially
reduces per-chunk segmentation noise at the cost of increased
computation.

\section{Block 2 --- WavLM Feature Extraction}
\label{sec:block2}

\paragraph{Concept:}
Each audio chunk is passed through the pretrained WavLM-Large encoder
to extract frame-level acoustic representations. Unlike conventional
systems that use a single handcrafted feature (e.g., MFCCs), DiariZen
leverages the rich representational hierarchy of a self-supervised
foundation model. All 25 internal representations, one from the CNN
feature extractor and 24 from the transformer layers, are combined
via a learned weighted sum, following the SUPERB
framework~\citep{yang2021superb}. This allows the model to learn which
layers are most informative for the diarization task.

\paragraph{Source code:} This block corresponds to \texttt{Model.forward()} in
\texttt{diarizen/models/eend/model\_wavlm\_conformer.py},
specifically the \texttt{wav2wavlm()} method that calls
\texttt{wavlm\_model.extract\_features()}, followed by
\texttt{self.weight\_sum()} which applies the learned scalar combination.

\paragraph{Implementation:}
The pruned WavLM-Large CNN feature extractor downsamples the input by
a factor of $5 \times 2^6 = 320$, converting 256,000 input samples into
$\lfloor (256{,}000 - 400) / 320 \rfloor + 1 = 799$ output frames at
approximately 50 frames per second. The transformer produces 1024-dimensional
representations per frame per layer. The full layer stack
$\mathbf{F} \in \mathbb{R}^{B \times T \times 1024 \times 25}$ is
combined as:
\begin{equation}
  \hat{\mathbf{f}} = \mathbf{F} \mathbf{w}, \quad
  \mathbf{w} \in \mathbb{R}^{25},
\end{equation}
where $\mathbf{w}$ is the weight vector learned by
\texttt{weight\_sum = nn.Linear(25, 1, bias=False)}, producing the
output tensor of shape \shape{10, 799, 1024}.

\paragraph{Structural pruning:}
The WavLM-Large model was pruned from 316M to 63M parameters (80\%
reduction) using DPHuBERT-style structured
pruning~\citep{peng2023dphubert} applied jointly with diarization
fine-tuning. The pruning removes 76.6\% of attention heads, 85.3\% of
FFN intermediate dimensions, and completely disables 4 of 24 transformer
layers. Crucially, the transformer embedding dimension (1024) remains
unchanged, so the interface with subsequent blocks is unaffected.

\begin{figure}[h]
  \centering
  \includegraphics[width=0.95\linewidth]{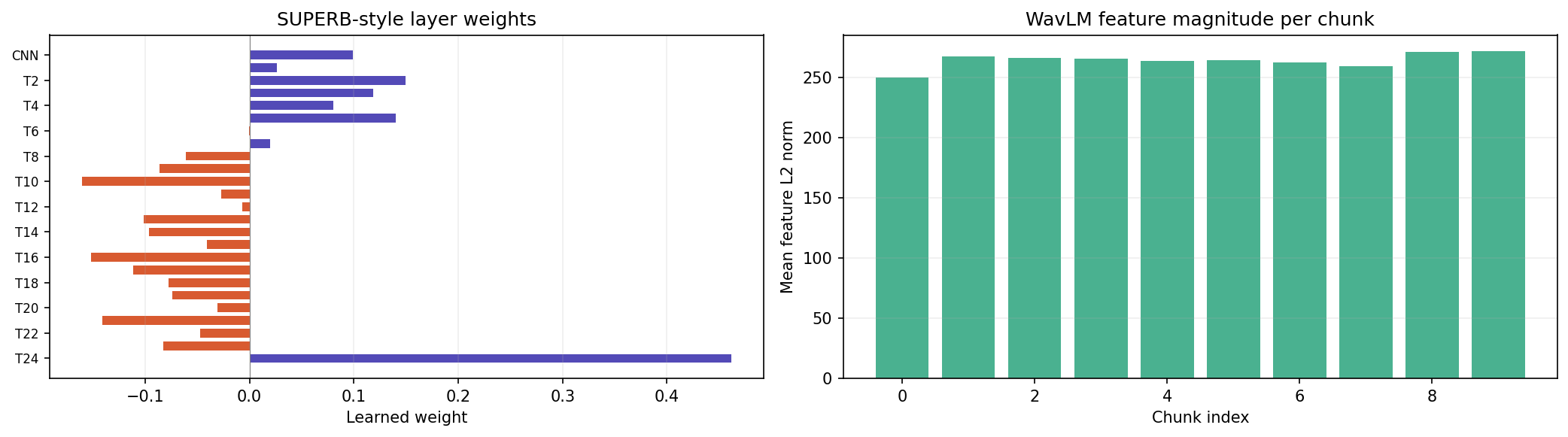}
  \caption{Left: Learned SUPERB-style layer weights for the 25 WavLM
    layers (blue = positive, orange = negative). Early transformer layers
    and the final layer receive the largest positive weights, consistent
    with speaker identity being primarily encoded in early and summary
    representations. Right: Mean feature L2 norm per chunk, showing
    stable representations across the 10 sliding windows.}
  \label{fig:block2}
\end{figure}

\paragraph{Visualization:}
Figure~\ref{fig:block2} shows the learned layer weights and feature
magnitudes. Early transformer layers (T2--T5) and the final layer (T24)
receive the largest positive weights, while middle layers (T8--T22)
receive negative weights. This pattern suggests that speaker identity
information is concentrated in early acoustic layers and the final
summary layer, while middle layers encoding linguistic content are
actively suppressed~\citep{yang2021superb, chen2022wavlm}.

\section{Block 3 --- Conformer Backend and Powerset Classification}
\label{sec:block3}

\paragraph{Concept:}
The WavLM features are passed through a lightweight Conformer
encoder followed by a linear classifier that predicts, for each frame,
which \emph{combination} of speakers is active. The Conformer models
temporal context that the sliding WavLM encoder cannot capture alone,
and the powerset head enables confident multi-speaker predictions with
clear overlap modelling.

\paragraph{Source code:}
This block corresponds to the remainder of \texttt{Model.forward()} in
\texttt{model\_wavlm\_conformer.py}, calling in
sequence: \texttt{self.proj}, \texttt{self.lnorm},
\texttt{self.conformer}, \texttt{self.classifier}, and
\texttt{self.activation}.

\paragraph{Implementation:}
The data flow through \block{3} is:
\begin{equation}
\begin{aligned}
  \underbrace{\shape{10,799,1024}}_{\text{WavLM}}
  &\xrightarrow{\;\text{proj}\;} \shape{10,799,256}
   \xrightarrow{\;\text{lnorm}\;} \shape{10,799,256}
   \xrightarrow{\;4\times\text{Conformer}\;} \shape{10,799,256} \\[6pt]
  &\xrightarrow{\;\text{classifier}\;} \shape{10,799,11}
   \xrightarrow{\;\text{LogSoftmax}\;} \shape{10,799,11}.
\end{aligned}
\end{equation}

Each \texttt{ConformerBlock} consists of four sub-modules applied
sequentially with residual connections: a scaled feed-forward network
(FFN, $256 \rightarrow 1024 \rightarrow 256$, Macaron-style), a
multi-head self-attention module (4 heads, $d_k=64$, no relative
positional encoding), a depthwise convolutional module (kernel size 31,
$\approx$620\,ms receptive field at 50\,fps), and a second FFN followed
by layer normalization. The \texttt{activation} is
\texttt{LogSoftmax(dim=-1)}, consistent with the powerset log-likelihood
objective used during training.

\paragraph{Powerset to multilabel conversion:}
The raw model output consists of log-probabilities over 11 powerset
classes. To obtain per-speaker binary activity, \texttt{Powerset.to\_multilabel()}
maps each frame's argmax class to the corresponding binary vector via:
\begin{equation}
  \mathbf{m} = \text{one\_hot}(\arg\max \hat{\mathbf{p}}) \cdot
  \mathbf{M},
\end{equation}
where $\mathbf{M} \in \{0,1\}^{11 \times 4}$ is the fixed powerset
mapping matrix. The output multilabel tensor has shape \shape{10,799,4}.

\begin{figure}[h]
  \centering
  \includegraphics[width=0.95\linewidth]{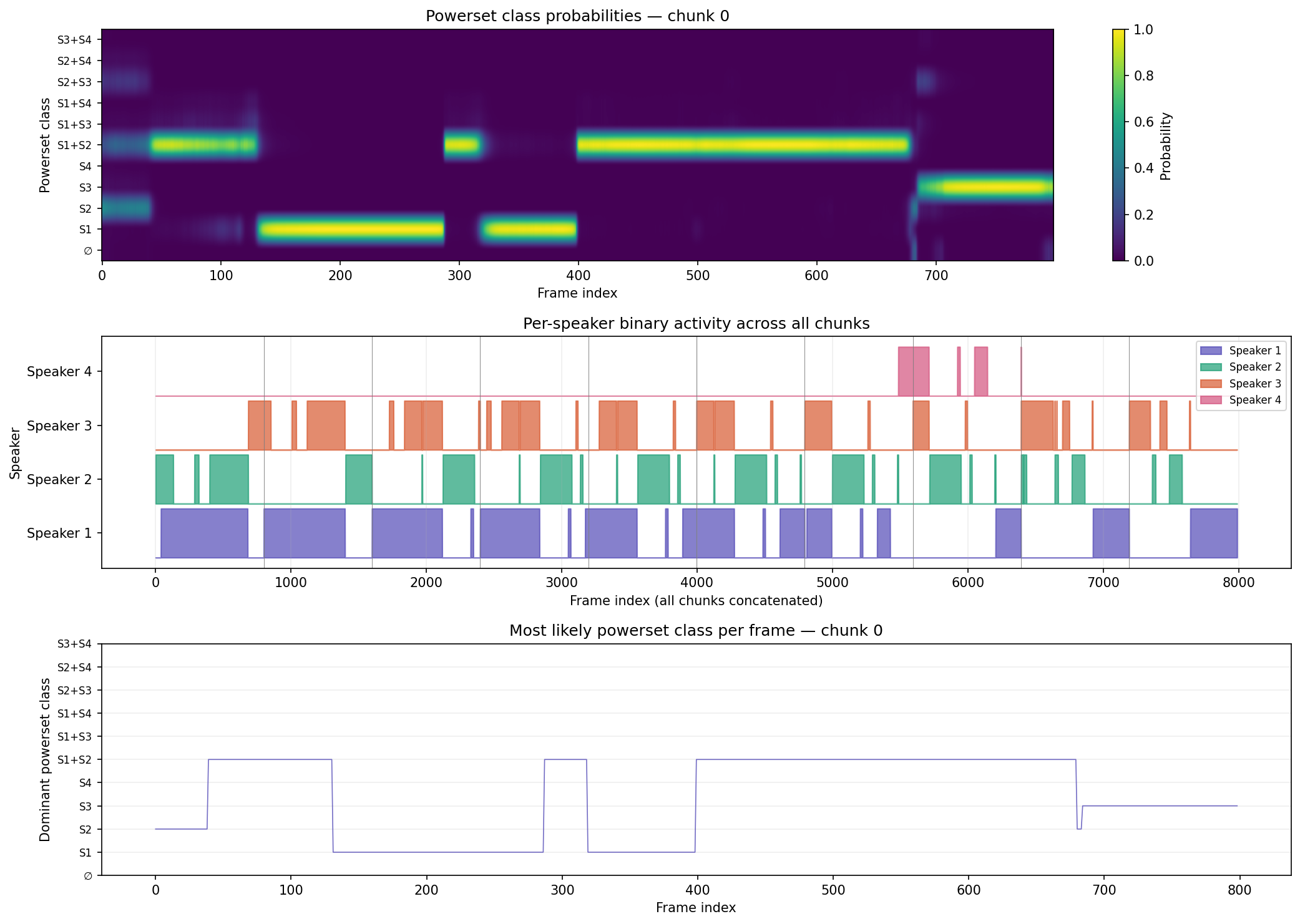}
\caption{Block~3 outputs for \texttt{EN2002a\_30s.wav}.
  Top: Powerset class probability heatmap for chunk~0. Yellow
  = high probability ($\approx$1.0), purple = near zero. The
  dominant classes are class~5 (two-speaker overlap) and class~1
  (single speaker), with sharp confident transitions between states.
  Middle: Per-speaker binary activity across all 10 chunks
  concatenated (7,990 frames = 30 seconds), derived via
  \texttt{Powerset.to\_multilabel()}.
  Bottom: Most likely powerset class per frame for chunk~0,
  showing the argmax class index over time and confirming that the
  model commits to a single class per frame with high confidence.}
  \label{fig:block3}
\end{figure}

\paragraph{Visualization:}
Figure~\ref{fig:block3} shows the powerset class probability heatmap
for chunk~0 and the per-speaker activity across all 10 chunks. The sharp
yellow bands in the heatmap confirm high model confidence, with class~5
(two-speaker overlap, $\{$Speaker\,1, Speaker\,2$\}$) dominating the
first 13 seconds of the recording.

\section{Block 4 --- Segmentation Aggregation via Overlap-Add}
\label{sec:block4}

\paragraph{Concept:}
Each of the 10 chunks produces an independent per-frame speaker activity
prediction. Since consecutive chunks overlap by 90\%, most frames of
the recording are covered by up to 10 independent predictions. This
block averages all predictions covering each output frame
(overlap-add, OLA), applies a temporal median filter to remove brief
spurious activations, and estimates the instantaneous number of active
speakers per frame.

\paragraph{Source code:}
This block corresponds to \texttt{DiariZenPipeline.\_\_call\_\_()} in \texttt{inference.py}, calling
\texttt{self.get\_segmentations()}, followed by
\texttt{scipy.ndimage.median\_filter()} and
\texttt{self.speaker\_count()}.

\paragraph{Implementation:}
\texttt{get\_segmentations()} internally calls
\texttt{pyannote.audio.core.\allowbreak Inference.\_\_aggregate\_\_()} which
computes the OLA average:
\begin{equation}
  \hat{\mathbf{s}}[t] =
  \frac{\sum_{c: t \in \mathcal{W}_c} \mathbf{s}_c[t - \tau_c]}
       {\sum_{c: t \in \mathcal{W}_c} 1},
\end{equation}
where $\mathcal{W}_c$ is the time interval covered by chunk $c$ and
$\tau_c$ is its start frame offset. The median filter is applied with
kernel size $(1, 11, 1)$, smoothing over 11 frames ($\approx$220\,ms)
along the time axis while leaving the chunk and class dimensions
unchanged. The speaker count is then estimated as the sum of active
speakers per output frame, clipped to \texttt{max\_speakers=20}.

\paragraph{Statistics on the test recording:}
The speaker count distribution over the 1,521 output frames of
\texttt{EN2002a\_30s.wav} is: 8.0\% silence (0 speakers), 64.0\%
single-speaker (1 speaker), and 27.9\% overlap (2 speakers). The
maximum instantaneous count is 2, consistent with
\texttt{max\_speakers\_per\_frame=2}.

\begin{figure}[H]
  \centering
  \includegraphics[width=0.95\linewidth]{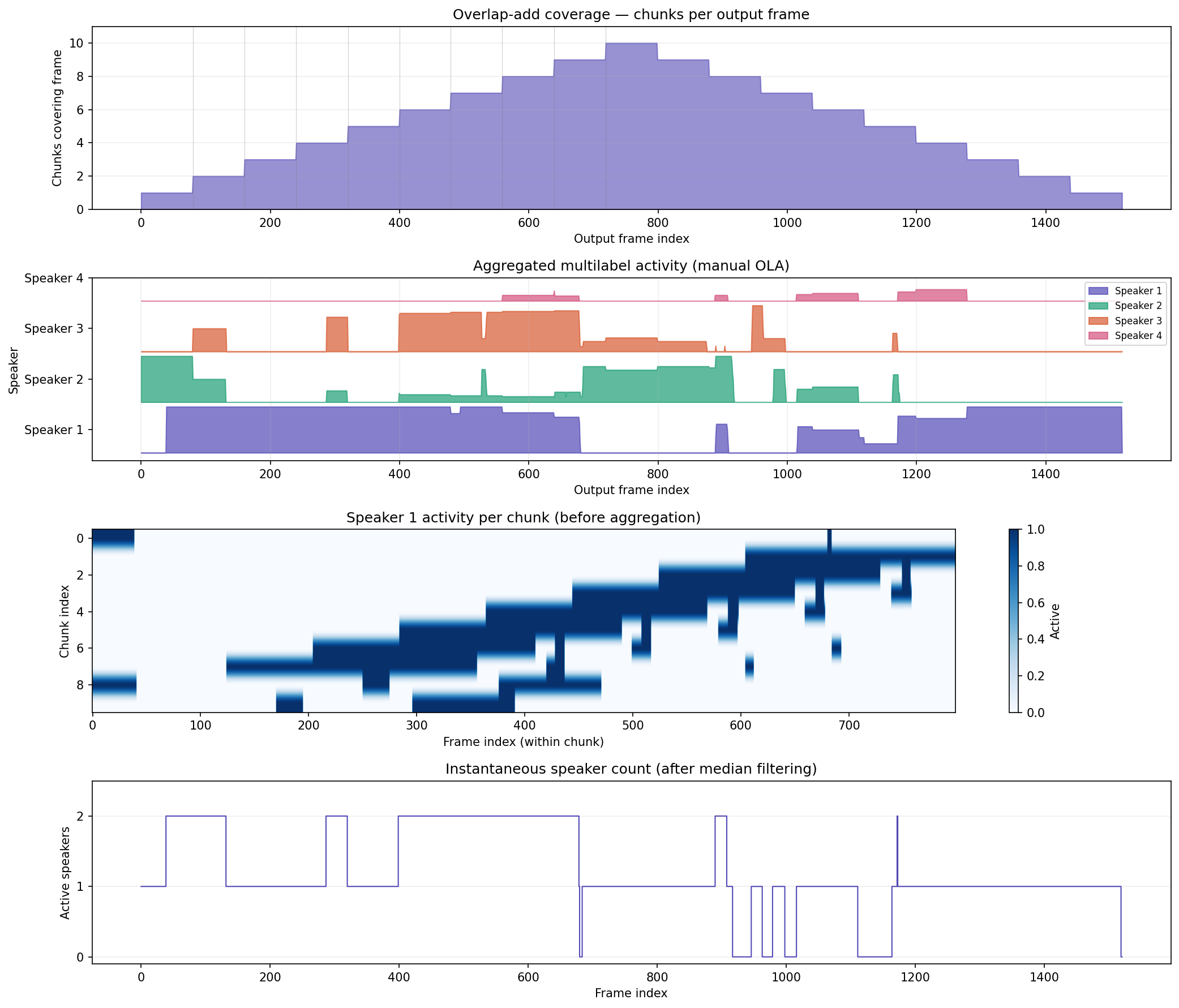}
\caption{Block~4 outputs for \texttt{EN2002a\_30s.wav}.
  \textbf{Top:} Overlap-add coverage map — number of chunks
  covering each output frame (up to 10 for interior frames).
  \textbf{Second:} Aggregated per-speaker activity after
  overlap-add averaging and median filtering.
  \textbf{Third:} Per-chunk activity heatmap for local
  Speaker~1 before aggregation.
  \textbf{Bottom:} Instantaneous speaker count after median
  filtering (0 = silence, 1 = single speaker, 2 = overlap);
  8.0\% silence, 64.0\% single-speaker, 27.9\% overlap.}
  \label{fig:block4}
\end{figure}

\paragraph{Visualization:}
Figure~\ref{fig:block4} shows the instantaneous speaker count over the
full 30\,s recording after overlap-add aggregation and median filtering.
Three distinct conversational phases are visible: heavy overlap in the
first half (0--13.6\,s), rapid turn-taking in the middle (13.6--23\,s),
and a single dominant speaker closing the recording (23--30\,s).

\section{Block 5 --- Speaker Embedding Extraction}
\label{sec:block5}

\paragraph{Concept:}
For each (chunk, local speaker) pair, a 256-dimensional speaker
embedding is extracted using the WeSpeaker ResNet34 model. The
embedding represents the speaker's voice characteristics in a compact
form suitable for clustering. A key design decision is
\emph{overlap exclusion}: frames where two or more speakers are
simultaneously active are masked out before embedding extraction,
ensuring that only clean single-speaker frames contribute to the
embedding.

\paragraph{Source code:}
This block corresponds to
\texttt{SpeakerDiarization.get\_embeddings()} in
\texttt{pyannote-audio/pyannote/audio/pipelines/speaker\_diarization.py}
, specifically the \texttt{iter\_waveform\_and\_mask()}
inner function and the batched call to \texttt{self.\_embedding()}.

\paragraph{Implementation:}
For each chunk $c$ and local speaker $k$, the clean mask is:
\begin{equation}
  \mathbf{m}_{c,k}[t] =
  \mathbf{s}_{c,k}[t] \cdot
  \mathbf{1}\!\left[\sum_{k'} \mathbf{s}_{c,k'}[t] < 2\right],
\end{equation}
where the indicator function zeroes out frames with 2 or more simultaneous
speakers. The WeSpeaker model applies masked weighted average pooling
over the clean frames to produce a 256-dim embedding, which is
L2-normalised so that cosine similarity equals the dot product. If the
number of clean frames falls below a minimum threshold
\texttt{min\_num\_frames}, the full mask (including overlap frames) is
used as a fallback.

In the test recording, all 40 (chunk, speaker) pairs produce valid
embeddings (no \texttt{NaN} values), indicating that every local speaker
has sufficient clean speech in every chunk.

\begin{figure}[H]
  \centering
  \includegraphics[width=0.95\linewidth]{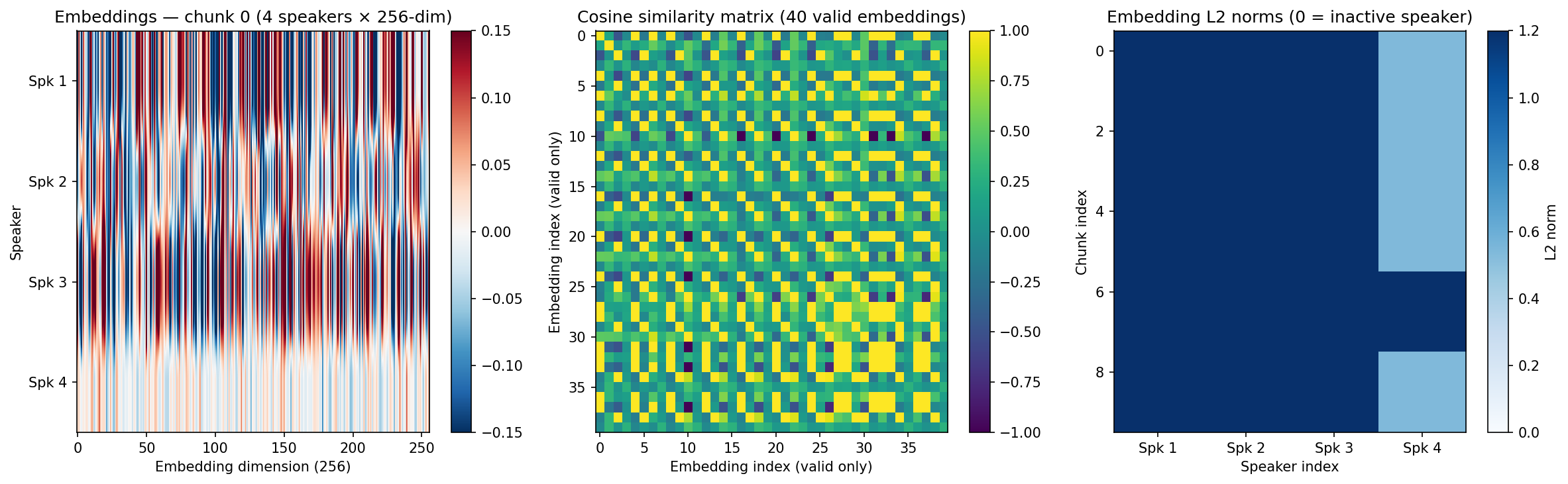}
\caption{Block~5 outputs for \texttt{EN2002a\_30s.wav}.
  Left: Raw embedding values for chunk~0 (4 speakers
  $\times$ 256-dim), showing the high-dimensional
  speaker representation before L2 normalisation.
  Middle: Pairwise cosine similarity matrix over all 40
  valid embeddings. Yellow cells ($\approx$1.0) indicate
  highly similar pairs, likely belonging to the same global
  speaker. The scattered yellow pattern reflects the
  cross-chunk same-speaker structure that VBx exploits in
  \block{6}.
  Right: Embedding L2 norms for all 40 (chunk, speaker)
  pairs. All norms are close to 1.0 due to L2 normalisation;
  the light-blue cells for local Speaker~4 in several chunks
  indicate low-confidence embeddings from insufficient clean
  frames.}
  \label{fig:block5}
\end{figure}

\paragraph{Visualization:}
Figure~\ref{fig:block5} shows the embedding L2 norms and cosine
similarity matrix for the 40 (chunk, speaker) pairs. All embeddings
have norm $\approx 1.0$ since they are L2-normalised. The scattered
yellow squares in the similarity matrix correspond to same-speaker
pairs across chunks — the cross-chunk similarity structure that
VBx will exploit in \block{6}.

\section{Block 6 --- VBx Clustering}
\label{sec:block6}

\paragraph{Concept:}
The 40 speaker embeddings (10 chunks $\times$ 4 local speakers) must be
grouped into a small number of global speaker identities. This is the
step that resolves the \emph{local permutation problem}: the Conformer
assigns local speaker IDs (0, 1, 2, 3) independently per chunk with no
consistency constraint, so local Speaker~0 in Chunk~0 may refer to a
completely different person than local Speaker~0 in Chunk~5. VBx
clustering finds the globally consistent assignment.

\paragraph{Source code:}
This block corresponds to \texttt{DiariZenPipeline.\_\_call\_\_()} in \texttt{inference.py}, calling \texttt{self.clustering()}.
The VBx implementation is part of the \texttt{diarizen} package and
uses PLDA model files loaded from \texttt{diarizen\_hub/plda/}.

\paragraph{Implementation:}
VBx operates in two stages:
\begin{enumerate}
  \item \textbf{AHC with PLDA.} The embeddings are projected from
    256-dim to 128-dim via LDA. Pairwise PLDA log-likelihood scores are
    computed and used as affinities for agglomerative hierarchical
    clustering with \texttt{ahc\_threshold=0.6}.
  \item \textbf{VB-HMM refinement.} Starting from the AHC initialisation,
    a Variational Bayes EM algorithm over a hidden Markov model refines
    speaker assignments iteratively (\texttt{max\_iters=20},
    \texttt{Fa=0.07}, \texttt{Fb=0.8}).
\end{enumerate}

For the test recording, VBx correctly identifies 4 global speakers.
Of the 40 (chunk, speaker) pairs, 8 are assigned $-2$ (inactive),
meaning those local speaker slots had no active speech in those chunks.

\begin{figure}[h]
  \centering
  \includegraphics[width=0.95\linewidth]{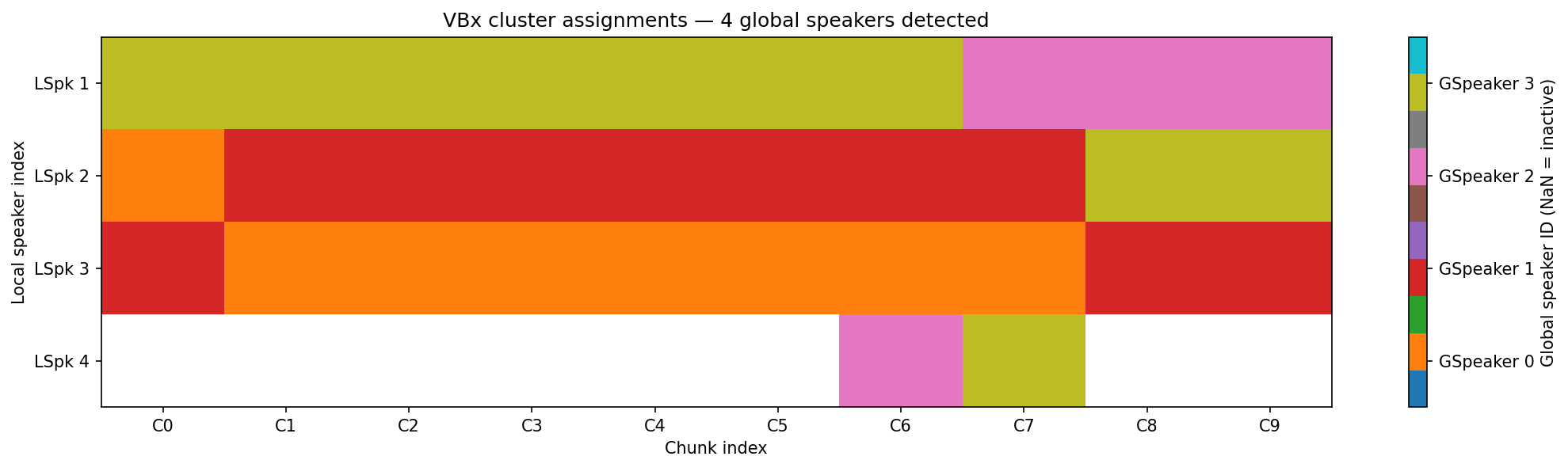}
  \caption{VBx cluster assignments: each cell shows the global speaker
    ID assigned to a (chunk, local speaker) pair. White cells indicate
    inactive speaker slots assigned as $-2$. Note how the same global speaker
    (colour) appears at different local indices across chunks,
    this is the local permutation ambiguity that VBx resolves.}
  \label{fig:block6}
\end{figure}

\paragraph{Visualisation:}
Figure~\ref{fig:block6} shows the cluster assignment matrix. The same
global speaker identity (colour) appears at different local speaker
indices across chunks, illustrating the permutation ambiguity that
VBx resolves.

\section{Block 7 --- Reconstruction and RTTM Output}
\label{sec:block7}

\paragraph{Concept:}
The final block converts the VBx cluster assignments into a
continuous diarization annotation and writes the output in the standard
rich transcription time mark (RTTM) format. This involves three steps:
(1) mapping local to global speaker identities and re-aggregating via
OLA, (2) binarising the continuous scores to produce discrete active
segments, and (3) renaming integer cluster IDs to human-readable labels.

\paragraph{Source code:}
This block corresponds to \texttt{DiariZenPipeline.\_\_call\_\_()} in \texttt{inference.py}, calling \texttt{self.reconstruct()},
\texttt{Binarize(onset=0.5)}, \texttt{result.rename\_labels()}, and
\texttt{result.to\_rttm()}.

\paragraph{Implementation:}
\texttt{reconstruct()} builds a clustered segmentation tensor:
\begin{equation}
  \hat{\mathbf{s}}_{c,t,k} =
  \max_{\ell:\, \text{cluster}(c,\ell)=k} \mathbf{s}_{c,t,\ell},
\end{equation}
for each global speaker $k$, taking the maximum activity score over all
local speakers assigned to $k$ in chunk $c$. The resulting
\texttt{SlidingWindowFeature} of shape \shape{1521, 4} is passed to
\texttt{to\_diarization()}, which applies a second OLA pass to obtain
the continuous global-speaker activity timeline. \texttt{Binarize} with
onset and offset both set to 0.5 then identifies contiguous active
regions for each speaker, producing the final pyannote
\texttt{Annotation} object.

\paragraph{Results on EN2002a\_30s.wav.}
The pipeline produces 4 speakers and 13 segments. The dominant speaker
(\texttt{SPEAKER\_03}) holds the floor for 12.8~s continuously (0.79\,s
to 13.61\,s), consistent with the meeting discussion style of the AMI
corpus. \texttt{SPEAKER\_02} takes over for the final 6.9\,s
(23.45\,s to 30.39\,s). The complete RTTM is:

\begin{lstlisting}[language={}, caption={RTTM output for EN2002a\_30s.wav.}]
SPEAKER EN2002a_30s 1  0.013  2.640 <NA> <NA> SPEAKER_00 <NA> <NA>
SPEAKER EN2002a_30s 1  0.792 12.820 <NA> <NA> SPEAKER_03 <NA> <NA>
SPEAKER EN2002a_30s 1  5.753  0.660 <NA> <NA> SPEAKER_00 <NA> <NA>
SPEAKER EN2002a_30s 1  7.993  2.560 <NA> <NA> SPEAKER_00 <NA> <NA>
SPEAKER EN2002a_30s 1 10.553  0.140 <NA> <NA> SPEAKER_01 <NA> <NA>
SPEAKER EN2002a_30s 1 10.692  2.900 <NA> <NA> SPEAKER_00 <NA> <NA>
SPEAKER EN2002a_30s 1 13.692  4.660 <NA> <NA> SPEAKER_01 <NA> <NA>
SPEAKER EN2002a_30s 1 17.812  0.360 <NA> <NA> SPEAKER_03 <NA> <NA>
SPEAKER EN2002a_30s 1 18.933  0.340 <NA> <NA> SPEAKER_00 <NA> <NA>
SPEAKER EN2002a_30s 1 19.593  0.380 <NA> <NA> SPEAKER_01 <NA> <NA>
SPEAKER EN2002a_30s 1 20.332  1.900 <NA> <NA> SPEAKER_03 <NA> <NA>
SPEAKER EN2002a_30s 1 23.293  0.180 <NA> <NA> SPEAKER_01 <NA> <NA>
SPEAKER EN2002a_30s 1 23.453  6.940 <NA> <NA> SPEAKER_02 <NA> <NA>
\end{lstlisting}

\begin{figure}[h]
  \centering
  \includegraphics[width=0.95\linewidth]{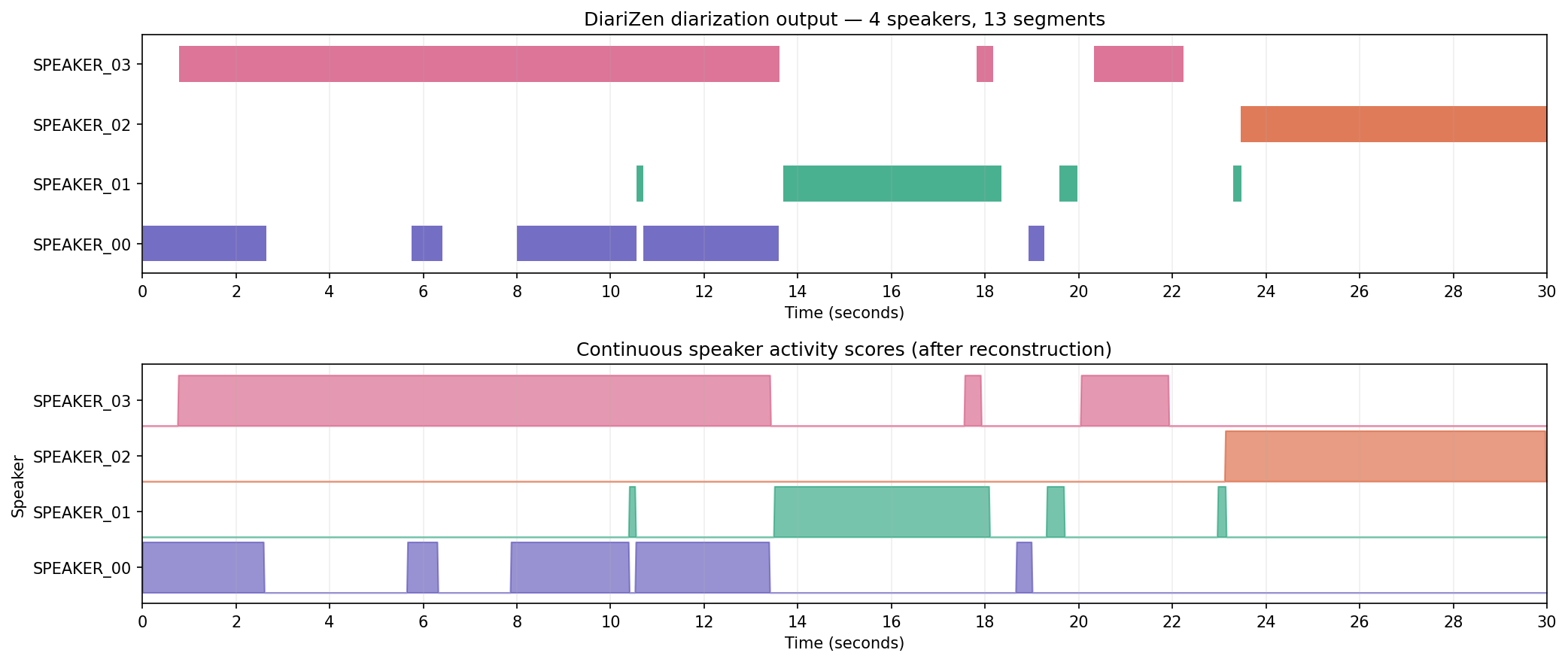}
  \caption{Final diarization output for \texttt{EN2002a\_30s.wav}: 4
    speakers, 13 segments over 30~seconds. Each bar represents one
    RTTM segment. \texttt{SPEAKER\_03} (pink) dominates the first half
    with a 12.8~s continuous turn. \texttt{SPEAKER\_02} (orange) closes
    the recording with a 6.9~s uninterrupted segment.}
  \label{fig:block7}
\end{figure}

\section{Experimental Demonstration}
\label{sec:experiment}

\subsection{Test recording}

We demonstrate the pipeline on \texttt{EN2002a\_30s.wav}, a 30-second
excerpt from session EN2002a of the AMI Meeting
Corpus~\citep{carletta2005ami}. This recording is a far-field
single-distant-microphone (SDM) capture of a multi-speaker meeting
containing 4 participants, with significant overlap (27.9\% of frames
have 2 simultaneously active speakers by the model's count).

\subsection{Results}

Table~\ref{tab:results} summarises the pipeline outputs. Full DER
evaluation against ground truth annotations requires the complete AMI
reference RTTM, which is available through the LDC~\citep{carletta2005ami}. The pipeline
completes the 30-second recording in under 60 seconds on a single H200
GPU, including model loading from cache.

\begin{table}[h]
  \caption{Pipeline output summary for \texttt{EN2002a\_30s.wav}.}
  \label{tab:results}
  \centering
  \begin{tabular}{ll}
    \toprule
    Metric & Value \\
    \midrule
    Audio duration            & 30.0~s \\
    Number of chunks          & 10 (9 complete + 1 zero-padded) \\
    Frames per chunk          & 799 (at $\approx$50~fps) \\
    Powerset classes          & 11 ($S=4$, $O=2$) \\
    Overlap frames            & 27.9\% \\
    Detected global speakers  & 4 \\
    Output segments           & 13 \\
    Longest segment           & 12.82~s (SPEAKER\_03) \\
    Shortest segment          & 0.14~s (SPEAKER\_01) \\
    \bottomrule
  \end{tabular}
\end{table}

\section{Conclusion}
\label{sec:conclusion}

We have presented a self-contained tutorial of the complete DiariZen
speaker diarization pipeline, decomposing it into seven functional
blocks with detailed explanations, source code references, and annotated
visualisations. Our accompanying GitHub repository provides standalone
executable scripts for each block and a Jupyter notebook for interactive
exploration, making the state-of-the-art DiariZen system accessible
to researchers and practitioners who wish to understand, reproduce, or
extend it. Future work will address the identified architectural gaps, with
particular focus on integrating adaptive affinity pruning from
SC-pNA~\citep{raghav2025scpna} into the VBx clustering stage and
evaluating the combined system on DIHARD-III~\citep{ryant21_interspeech} and the DISPLACE-M~\citep{e2026benchmarkingspeechsystemsfrontline}
challenge datasets.


\begin{ack}
The author thanks Dr.\ Md Sahidullah at TCG CREST for his guidance and valuable
feedback on this work, and the DiariZen team at Brno University of
Technology for the open-source release of their system. The
experiments were conducted on the H200 GPU cluster at TCG CREST.
\end{ack}

\bibliographystyle{plainnat}
\bibliography{references}

@inproceedings{han2025leveraging,
  title={Leveraging self-supervised learning for speaker diarization},
  author={Han, Jiangyu and Landini, Federico and Rohdin, Johan and Silnova, Anna and Diez, Mireia and Burget, Luk{\'a}{\v{s}}},
  booktitle={Proc. ICASSP},
  year={2025}
}

@article{han2025fine,
  title={Fine-tune Before Structured Pruning: Towards Compact and Accurate Self-Supervised Models for Speaker Diarization},
  author={Han, Jiangyu and Landini, Federico and Rohdin, Johan and Silnova, Anna and Diez, Mireia and Cernocky, Jan and Burget, Lukas},
  journal={arXiv preprint arXiv:2505.24111},
  year={2025}
}

@article{han2025efficient,
  title={Efficient and Generalizable Speaker Diarization via Structured Pruning of Self-Supervised Models},
  author={Han, Jiangyu and P{\'a}lka, Petr and Delcroix, Marc and Landini, Federico and Rohdin, Johan and Cernock{\`y}, Jan and Burget, Luk{\'a}{\v{s}}},
  journal={arXiv preprint arXiv:2506.18623},
  year={2025}
}

@inproceedings{bredin2023pyannote,
  author    = {Herv{\'{e}} Bredin},
  title     = {pyannote.audio 2.1 speaker diarization pipeline:
               Principle, benchmark, and recipe},
  booktitle = {Proc. INTERSPEECH},
  year      = {2023},
  pages     = {1983--1987},
}

@article{chen2022wavlm,
  author    = {Sanyuan Chen and Chengyi Wang and Zhengyang Chen and
               Yu Wu and Shujie Liu and Zhuo Chen and Jinyu Li and
               Naoyuki Kanda and Takuya Yoshioka and Xiong Xiao and
               Jian Wu and Long Zhou and Shuo Ren and Yanmin Qian and
               Yao Qian and Jian Wu and Michael Zeng and Furu Wei},
  title     = {{WavLM}: Large-Scale Self-Supervised Pre-Training for
               Full Stack Speech Processing},
  journal   = {IEEE Journal of Selected Topics in Signal Processing},
  volume    = {16},
  number    = {6},
  pages     = {1505--1518},
  year      = {2022},
}

@inproceedings{yang2021superb,
  author    = {Shu{-}wen Yang and Po{-}Han Chi and Yung{-}Sung Chuang and
               Cheng{-}I Jeff Lai and Kushal Lakhotia and Yist Y. Lin and
               Andy T. Liu and Jiatong Shi and Xuankai Chang and
               Guan{-}Ting Lin and Tzu{-}Hsien Huang and Wei{-}Cheng Tseng and
               Ko{-}tik Lee and Da{-}Rong Liu and Zili Huang and
               Shuyan Dong and Shang{-}Wen Li and Shinji Watanabe and
               Abdelrahman Mohamed and Hung{-}yi Lee},
  title     = {{SUPERB}: Speech Processing Universal PERformance Benchmark},
  booktitle = {Proc. INTERSPEECH},
  year      = {2021},
  pages     = {1194--1198},
}

@inproceedings{peng2023dphubert,
  author    = {Peng, Puyuan and Harwath, David},
  title     = {{DPHuBERT}: Joint Distillation and Pruning of
               Self-Supervised Speech Models},
  booktitle = {Proc. INTERSPEECH},
  year      = {2023},
  pages     = {3627--3631},
}

@inproceedings{plaquet2023powerset,
  author    = {Alexis Plaquet and Herv{\'{e}} Bredin},
  title     = {Powerset Multi-Class Cross Entropy Loss for Neural
               Speaker Diarization},
  booktitle = {Proc. INTERSPEECH},
  year      = {2023},
  pages     = {3353--3357},
}

@inproceedings{diez2019bayesian,
  author    = {Mireia D{\'{\i}}ez and Luk{\'{a}}{\v{s}} Burget and
               Federico Landini and Jan {\v{C}}ernock{\'{y}}},
  title     = {Analysis of {i-vector} Length Normalization in Speaker
               Recognition Systems},
  booktitle = {Proc. INTERSPEECH},
  year      = {2019},
  pages     = {2649--2653},
}

@inproceedings{prince2007probabilistic,
  author    = {Simon J. D. Prince and James H. Elder},
  title     = {Probabilistic Linear Discriminant Analysis for
               Inferences About Identity},
  booktitle = {Proc. IEEE International Conference on Computer Vision (ICCV)},
  year      = {2007},
  pages     = {1--8},
}

@inproceedings{fujita2019eend,
  author    = {Yusuke Fujita and Naoyuki Kanda and Shota Horiguchi and
               Kenji Nagamatsu and Shinji Watanabe},
  title     = {End-to-End Neural Speaker Diarization with
               Permutation-Free Objectives},
  booktitle = {Proc. INTERSPEECH},
  year      = {2019},
  pages     = {4300--4304},
}

@inproceedings{kinoshita2021integrating,
  author    = {Keisuke Kinoshita and Marc Delcroix and Shoko Araki and
               Tomohiro Nakatani},
  title     = {Integrating End-to-End Neural and Clustering-Based
               Diarization: Getting the Best of Both Worlds},
  booktitle = {Proc. IEEE International Conference on Acoustics, Speech
               and Signal Processing (ICASSP)},
  year      = {2021},
  pages     = {7198--7202},
}

@article{hsu2021hubert,
  author    = {Wei{-}Ning Hsu and Benjamin Bolte and
               Yao{-}Hung Hubert Tsai and Kushal Lakhotia and
               Ruslan Salakhutdinov and Abdelrahman Mohamed},
  title     = {{HuBERT}: Self-Supervised Speech Representation Learning
               by Masked Prediction of Hidden Units},
  journal   = {IEEE/ACM Transactions on Audio, Speech, and
               Language Processing},
  volume    = {29},
  pages     = {3451--3460},
  year      = {2021},
}

@article{park2022review,
  author    = {Tae Jin Park and Naoyuki Kanda and Dimitrios Dimitriadis and
               Kyu J. Han and Shinji Watanabe and Shrikanth Narayanan},
  title     = {A Review of Speaker Diarization: Recent Advances with
               Deep Learning},
  journal   = {Computer Speech \& Language},
  volume    = {72},
  pages     = {101317},
  year      = {2022},
}

@inproceedings{raghav2025scpna,
  author    = {Nikhil Raghav and Avisek Gupta and Md Sahidullah and
               Swagatam Das},
  title     = {Self-Tuning Spectral Clustering for Speaker Diarization},
  booktitle = {Proc. IEEE International Conference on Acoustics, Speech
               and Signal Processing (ICASSP)},
  year      = {2025},
  address   = {Hyderabad, India},
}

@inproceedings{ryant21_interspeech,
  title     = {The Third {DIHARD} Diarization Challenge},
  author    = {Neville Ryant and others},
  year      = {2021},
  booktitle = {Proc. INTERSPEECH}
}

@misc{e2026benchmarkingspeechsystemsfrontline,
      title={Benchmarking Speech Systems for Frontline Health Conversations: The DISPLACE-M Challenge}, 
      author={Dhanya E and Ankita Meena and Manas Nanivadekar and Noumida A and Victor Azad and Ashwini Nagaraj Shenoy and Pratik Roy Chowdhuri and Shobhit Banga and Vanshika Chhabra and Chitralekha Bhat and Shareef babu Kalluri and Srikanth Raj Chetupalli and Deepu Vijayasenan and Sriram Ganapathy},
      year={2026},
      eprint={2603.02813},
      archivePrefix={arXiv},
      primaryClass={eess.AS},
      url={https://arxiv.org/abs/2603.02813}, 
}

@inproceedings{carletta2005ami,
  title={The {AMI} meeting corpus: A pre-announcement},
  author={Carletta, Jean and others},
  booktitle={International Workshop on Machine Learning for Multimodal Interaction},
  pages={28--39},
  year={2005},
  organization={Springer}
}

\appendix
\section{Block Function Signatures}
\label{app:signatures}

The following summarizes the Python function signature for each block
module in the GitHub repository. Each function accepts the output
dictionary of the previous block and the loaded pipeline object,
and returns a new dictionary carrying all keys forward.

\begin{lstlisting}[caption={Block function signatures.}]
# Block 1 - no pipeline needed
b1 = run_block1(audio_path, diarizen_root, seg_duration=16.0,
                segmentation_step=0.1, batch_size=32)
# Blocks 2-7 - pipeline loaded once and passed through
pipeline = load_pipeline(diarizen_root, hf_cache)
b2 = run_block2(b1, pipeline)
b3 = run_block3(b2, pipeline)
b4 = run_block4(b3, pipeline)
b5 = run_block5(b4, pipeline)
b6 = run_block6(b5, pipeline)
b7 = run_block7(b6, pipeline, session_name="EN2002a_30s",
                rttm_out_dir="outputs/")
\end{lstlisting}

\end{document}